\documentclass[aps,prl,preprint,superscriptaddress,nofootinbib]{revtex4-2}
\usepackage[T1]{fontenc}
\usepackage{amsmath,amssymb,bm}
\usepackage{graphicx,xcolor,microtype}
\usepackage[colorlinks=true,linkcolor=teal!65!black,citecolor=teal!65!black,
 urlcolor=teal!65!black,breaklinks=true]{hyperref}
\usepackage{xurl}
\newcommand{\B}{\bm B}
\newcommand{\X}{\bm X}
\newcommand{\K}{\bm K}
\newcommand{\N}{\bm N}
\newcommand{\nn}{\bm n}
\newcommand{\grad}{\nabla}
\newcommand{\RR}{\mathbb R}
\DeclareMathOperator{\RePart}{Re}
\DeclareMathOperator{\ImPart}{Im}
\makeatletter
\def\frontmatter@abstractheading{}
\let\letter@abstract@produce\frontmatter@abstract@produce
\def\frontmatter@abstract@produce{%
 \begingroup\@booleanfalse\preprintsty@sw\letter@abstract@produce\endgroup}
\def\@author@present@script#1#2#3{%
 \begingroup
  \gdef\comma@space{}%
  \doauthor{#2}{#3}{\@affil@present@script}%
 \endgroup
 \advance\@tempcnta\m@ne
}
\def\frontmatter@thefootnote{\textdagger}
\def\frontmatter@footnote@produce{\global\let\@FMN@list\@empty}
\def\ps@titlepage{\let\@oddhead\@empty\let\@evenhead\@empty
 \def\@oddfoot{\hfil\footnotesize\textsuperscript{\textdagger}Email address for correspondence: lopezortizoe@ornl.gov\hfil}%
 \let\@evenfoot\@oddfoot}
\makeatother
\hypersetup{pdftitle={Exact nonsymmetric vacuum magnetic field with nested toroidal flux surfaces and nonzero rotational transform},
pdfsubject={Exact vacuum field, magnetic axis, nested toroidal flux surfaces, and nonzero transform}}

\begin{document}

\title{Exact nonsymmetric vacuum magnetic field with nested toroidal flux surfaces and nonzero rotational transform}
\author{Omar E. L\'opez}
\email{lopezortizoe@ornl.gov}
\affiliation{Fusion Energy Division, Oak Ridge National Laboratory
PO Box 2008, Oak Ridge, TN 37831, United States of America}
%\date{Oct 15, 2026}

\begin{abstract}
We show that an exact vacuum magnetic field without continuous Euclidean
symmetry can fill a bounded solid torus with nested
toroidal flux surfaces and nonzero rotational transform.  We construct such
a field in closed form and show that it is nowhere vanishing.  Every
off-axis field line closes after one toroidal transit and one poloidal turn,
with rotational transform $\iota=-1$ on every
flux surface.  The solution provides an exact benchmark for
three-dimensional equilibrium codes.
\end{abstract}

\maketitle

\section{Introduction}\label{sec:introduction}

A stellarator's external coils produce a nonaxisymmetric vacuum magnetic
field.  Nested toroidal magnetic surfaces are central to
magnetic-confinement fusion~\cite{helander2014}, but their existence is not
guaranteed without continuous symmetry: magnetic islands and stochastic
field line regions may occur.  Vacuum fields with nested surfaces are
therefore direct targets of coil optimization~\cite{giuliani2023} and can
serve as starting points for finite-pressure equilibrium
calculations~\cite{boozer2019curlfree}.
Numerical optimization can suppress islands and stochastic
regions~\cite{hansoncary1984}, but does not by itself establish the existence of an
exact field with nested surfaces throughout a toroidal volume.
Can a vacuum field without continuous Euclidean symmetry fill a bounded solid
torus with nested flux surfaces and nonzero rotational transform?  This
existence question has been open~\cite{cary1982,senguptaweitzner2019,landreman}.

Grad's longstanding conjecture held that toroidal equilibria with nested
pressure surfaces could form nontrivial smooth families only in the presence
of axial, helical, or plane-reflection symmetry~\cite{grad1967,grad1985,constantin2021}.
For sufficiently small pressure amplitude, plane-reflection-symmetric toroidal
equilibria with nonconstant pressure and zero rotational transform
exist~\cite{lortz1970}.  Nonsymmetric equilibria with
piecewise-constant pressure and current sheets were also
constructed~\cite{brunolaurence1996,enciso2025}.
Nontrivial smooth families of finite-pressure equilibria with none of these
symmetries were recently constructed, disproving this formulation of Grad's
conjecture~\cite{gomezserrano2026}.  Their pressure
surfaces are nested, their field lines close poloidally, and their magnetic
field vanishes on the axis.  A separate construction gives two further
counterexample families with the field, pressure, and nested flux surfaces
in closed form~\cite{landreman}.  One has $\iota=+2$ on every flux surface;
the other has a sheared transform profile.  Both families were subsequently
generalized to break stellarator symmetry~\cite{issan2026}.
These equilibria have nonconstant pressure and
therefore carry plasma current; they do not resolve the vacuum existence
question posed above.

A vacuum field satisfies $\grad\times\B=\bm0$ and
$\grad\mathbin{\cdot}\B=0$.  We seek such a field in a domain
$\Omega\subset\RR^3$ whose closure is a smooth solid torus, with
$\B\mathbin{\cdot}\bm{\nu}=0$ on the boundary $\partial\Omega$, where $\bm{\nu}$ is the
outward unit normal.  Perturbative approaches to three-dimensional vacuum
fields with dense nested flux surfaces were assessed through
surface-of-section calculations~\cite{cary1982,cary1984}.
Formal expansions to all orders were constructed for vacuum fields with
flux surfaces in a slab periodic in two directions~\cite{senguptaweitzner2019}.
Discrete reflection symmetry yields exact closed-line vacuum fields without
continuous symmetry in the same doubly periodic slab~\cite{weitzner}.  These constructions
use a flat periodic geometry and do not establish the corresponding result
in a solid torus embedded in $\RR^3$.
The vacuum field underlying an explicit construction of anisotropic
magnetohydrodynamic equilibria~\cite{satoyamada2023} is tangent to nested
tori in a domain without continuous Euclidean symmetry; the field itself is translation invariant and
has zero rotational transform.

Here we explicitly construct a vacuum field with nested toroidal flux
surfaces and rotational transform $\iota=-1$.  The field is nowhere
vanishing in $\Omega$, has a regular magnetic axis, and has no
continuous Euclidean symmetry.  Every off-axis field line closes after one
toroidal transit and one poloidal turn in the negative sense.
We select a nonzero complex-valued root $\zeta$ of an explicit
cubic polynomial from a Weierstrass representation~\cite{bairdwood1988,bairdwood1997}
and set $\B=\ImPart\,\grad\log\zeta$, yielding closed-form expressions for
the field and its first integrals.

\section{Explicit vacuum construction and global structure}
\label{sec:construction}
\label{sec:main-properties}

\subsection{Vacuum field and coordinate map}
\label{subsec:construction-field}

% Allow a page break before this display so Sec. II need not start a new page.
\begingroup\predisplaypenalty=0
For $\bm x=(x,y,z)\in\RR^3$, set $w=x+iy$ and $\bar w=x-iy$, and consider a
cubic polynomial obtained from the three-dimensional inhomogeneous Weierstrass
representation~\cite{bairdwood1988,bairdwood1997}:
\begin{align}
 G(\zeta;\bm x)
 &=2i\zeta^3+(\bar w+2)\zeta^2-2z\zeta-w,
 \label{eq:cubic}\\
 D(\zeta;\bm x)&=\partial_\zeta G
 =6i\zeta^2+2(\bar w+2)\zeta-2z.
 \label{eq:D}
\end{align}
\endgroup
Every nonzero simple root of Eq.~\eqref{eq:cubic} determines, by the
implicit-function theorem, a local real-analytic branch $\zeta(\bm x)$.  For
any such branch, set
\begin{equation}
 \begin{aligned}
 \N(\zeta)&=(1-\zeta^2,\ i(1+\zeta^2),\ 2\zeta),\\
 \B(\bm x)&=\ImPart\,\grad\log\zeta(\bm x)
 =\ImPart\!\left(\frac{\N(\zeta(\bm x))}
 {\zeta(\bm x)D(\zeta(\bm x);\bm x)}\right).
 \end{aligned}
 \label{eq:B}
\end{equation}
The spatial dependence of $\N$ is through $\zeta(\bm x)$, whereas $D$ also
depends explicitly on $\bm x$.  We suppress these arguments below.
Holding $\zeta$ fixed in $G$ and $D$ gives
$\grad_{\bm x}G=-\N$ and
$\bm M=\grad_{\bm x}D=(2\zeta,-2i\zeta,-2)$.  For
$\bm a,\bm b\in\mathbb C^3$,
$\bm a\mathbin{\cdot}\bm b=\sum_{j=1}^3a_jb_j$, with no complex
conjugation.  With
$\N'=d\N/d\zeta$, direct expansion gives
$\N\mathbin{\cdot}\N=\N'\mathbin{\cdot}\N
=\N\mathbin{\cdot}\bm M=0$.  Implicit differentiation and the chain rule
then give
\begin{equation}
 \begin{aligned}
 \grad\zeta&=\frac{\N}{D},\qquad
 \grad\zeta\mathbin{\cdot}\grad\zeta
 =\frac{\N\mathbin{\cdot}\N}{D^2}=0,\\
 \Delta\zeta
 &=\frac{\N'\mathbin{\cdot}\N-\N\mathbin{\cdot}\bm M}{D^2}-\frac{(\partial_\zeta D)\N\mathbin{\cdot}\N}{D^3}=0.
 \end{aligned}
 \label{eq:main-harmonic-chain}
\end{equation}
Consequently,
\begin{equation}
 \Delta\log\zeta
 =\frac{\Delta\zeta}{\zeta}
  -\frac{\grad\zeta\mathbin{\cdot}\grad\zeta}{\zeta^2}=0.
\end{equation}
Thus $\B$ is the gradient of the local harmonic phase
$\ImPart\log\zeta$, and
\begin{equation}
 \grad\times\B=\bm0,\qquad \grad\mathbin{\cdot}\B=0.
 \label{eq:main-curl-div}
\end{equation}

\begin{figure*}[t]
\centering
\includegraphics[width=\textwidth]{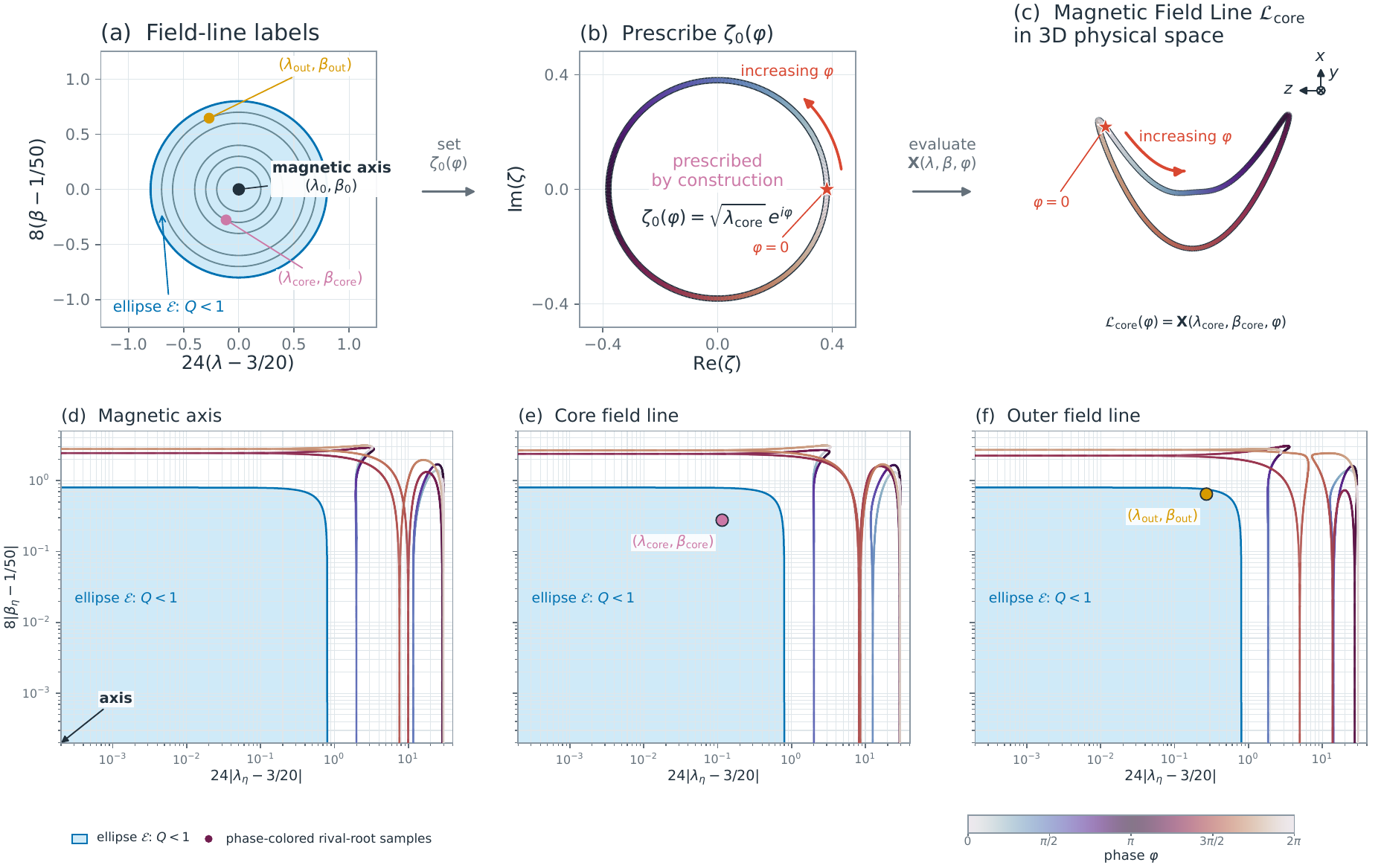}
\caption{From field line labels to a closed magnetic field line and
rival root separation.
(a) The open ellipse $\mathcal E$, its center, and several levels of $Q$.
The marked labels $(\lambda_{\mathrm{core}},\beta_{\mathrm{core}})$ and
$(\lambda_{\mathrm{out}},\beta_{\mathrm{out}})$ select the magnetic field lines
$\mathcal L_{\mathrm{core}}$ and $\mathcal L_{\mathrm{out}}$, respectively.
(b) The prescribed circle
$\zeta_0(\varphi)=\sqrt{\lambda_{\mathrm{core}}}e^{i\varphi}$ for the labels
$(\lambda_{\mathrm{core}},\beta_{\mathrm{core}})$ associated
with $\mathcal L_{\mathrm{core}}$.
(c) The resulting closed magnetic field line $\mathcal L_{\mathrm{core}}$.
In panels (b)--(f), color gives the phase $\varphi$.
(d)--(f) Absolute deviations $24|\lambda_\eta-\lambda_0|$ and
$8|\beta_\eta-\beta_0|$ of the labels associated with the two rival roots along
the magnetic axis, $\mathcal L_{\mathrm{core}}$, and
$\mathcal L_{\mathrm{out}}$, on logarithmic axes.  Solid points mark the
prescribed labels.  In panel (d), the arrow points toward the off-scale
magnetic axis labels at $(0,0)$.
These curves illustrate the rival root separation of
Eq.~\eqref{eq:main-rival-exclusion}.}
\label{fig:construction-flow}
\end{figure*}

We now introduce two real parameters $\lambda$ and $\beta$, which we prove in
Sec.~\ref{subsec:main-vacuum} are field line labels, and use them to identify
a bounded open domain on which exactly one of the three roots of
Eq.~\eqref{eq:cubic} is selected, giving a single global branch
$\zeta(\bm x)$.  The polynomial and
the domain of field line labels provide freedom to generate families of
vacuum configurations, as explored in
Supplemental Section~\ref{sec:supp-polynomial-program}.  For the explicit
configuration considered here, choose the center
\begin{equation}
 (\lambda_0,\beta_0)=\left(\frac3{20},\frac1{50}\right)
 \label{eq:main-parameter-center}
\end{equation}
and define the open ellipse $\mathcal E$ by
\begin{subequations}
\label{eq:main-parameter-sets}
\begin{align}
 Q(\lambda,\beta)
 &=900(\lambda-\lambda_0)^2+100(\beta-\beta_0)^2,
 \label{eq:main-q-tor}\\
 \mathcal E&=\{(\lambda,\beta):Q(\lambda,\beta)<1\}.
 \label{eq:main-torus-parameter-ellipse}
\end{align}
\end{subequations}
This particular ellipse permits a computer-assisted proof of global root
selection and injectivity
(Supplemental Section~\ref{sec:supp-interval-certificate}).
Let $\overline{\mathcal E}$ denote the closure of $\mathcal E$.  On
$\overline{\mathcal E}$, $\lambda\geq7/60>0$, so the prescribed root below
is nonzero.  Figure~\ref{fig:construction-flow}(a) illustrates $\mathcal E$,
its center, several levels of $Q$, and the core and outer field line labels
\begin{equation}
 (\lambda_{\mathrm{core}},\beta_{\mathrm{core}})
 =\left(\frac{151}{1040},-\frac{19}{1300}\right),\qquad
 (\lambda_{\mathrm{out}},\beta_{\mathrm{out}})
 =\left(\frac{433}{3120},\frac{131}{1300}\right),
 \label{eq:main-sample-lines}
\end{equation}
which lie on $Q=9/64$ and $49/64$, respectively.

Let $S^1=\RR/(2\pi\mathbb Z)$, identified with the unit complex numbers by
$\varphi\mapsto e^{i\varphi}$.  For
$(\lambda,\beta,\varphi)\in(0,\infty)\times\RR\times S^1$, prescribe
\begin{equation}
 \zeta_0=\sqrt\lambda e^{i\varphi}=u+iv.
 \label{eq:main-prescribed-root}
\end{equation}
At fixed $(\lambda,\beta)$, $\zeta_0(\varphi)$ traces a circle in the complex plane as $\varphi$
traverses $S^1$, as shown in Fig.~\ref{fig:construction-flow}(b) for the magnetic field line with labels $(\lambda_{\mathrm{core}},\beta_{\mathrm{core}})$.
For the prescribed values $(\lambda,\beta,\varphi)$, we determine
$\bm x$ by requiring $G(\zeta_0;\bm x)=0$ together with
\begin{equation}
 \beta=\nn\mathbin{\cdot}\bm x+\kappa,
 \label{eq:R}
\end{equation}
where
\begin{equation}
\begin{aligned}
 \bm A&=\RePart\N(\zeta_0),\qquad \bm C=\ImPart\N(\zeta_0),\qquad
 H=2\zeta_0^2+2i\zeta_0^3,\\
 \nn&=\frac{(2u,2v,\lambda-1)}{1+\lambda},\qquad
 \kappa=\frac{4\lambda u(1-v)}{(1+\lambda)^2}.
\end{aligned}
\label{eq:main-frame-data}
\end{equation}
Direct calculation shows that $\bm A$, $\bm C$, and $\nn$ are mutually
orthogonal, with $|\bm A|^2=|\bm C|^2=(1+\lambda)^2$ and $|\nn|^2=1$; the
matrix with these vectors as rows has determinant $-(1+\lambda)^2$.  Hence the
real and imaginary parts of $G(\zeta_0;\bm x)=0$, together with
Eq.~\eqref{eq:R}, form a nonsingular linear system for $(x,y,z)$.  Solving it
gives the explicit coordinate map
\begin{equation}
 \X(\lambda,\beta,\varphi)
 =\frac{\RePart(H)\bm A+\ImPart(H)\bm C}{(1+\lambda)^2}
 +(\beta-\kappa)\nn.
 \label{eq:main-forward}
\end{equation}
Differentiating Eq.~\eqref{eq:main-forward} gives
\begin{equation}
 \partial_\beta\X=\nn,\qquad |\nn|=1,\qquad
 \nn\mathbin{\cdot}\partial_\varphi\X=0.
 \label{eq:main-coordinate-normal}
\end{equation}
By construction,
$G\bigl(\zeta_0;\X(\lambda,\beta,\varphi)\bigr)=0$, so $\zeta_0$ is a root of
the cubic polynomial at $\X(\lambda,\beta,\varphi)$.  For each fixed $(\lambda,\beta)$,
the map $\varphi\mapsto\X(\lambda,\beta,\varphi)$ traces a closed magnetic
field line, as shown below. At the reference labels, write
$\mathcal L_{\mathrm{core}}(\varphi)=\X(\lambda_{\mathrm{core}},\beta_{\mathrm{core}},\varphi)$
and
$\mathcal L_{\mathrm{out}}(\varphi)=\X(\lambda_{\mathrm{out}},\beta_{\mathrm{out}},\varphi)$.
Figure~\ref{fig:construction-flow}(c) shows $\mathcal L_{\mathrm{core}}$, the
image of the circle in Fig.~\ref{fig:construction-flow}(b) under $\X$.

\subsection{Global coordinates and root selection}
\label{subsec:construction-domain}

Let $\eta=a+ib$ be any root of $G(\eta;\bm x)=0$, and set
$\lambda_\eta=|\eta|^2$.  We define $\beta_\eta(\bm x)$ by evaluating the
right-hand side of Eq.~\eqref{eq:R} at
$(u,v,\lambda)=(a,b,\lambda_\eta)$.  Thus every root $\eta$ at $\bm x$
has associated labels $(\lambda_\eta,\beta_\eta(\bm x))$.  At
$\bm x=\X(\lambda,\beta,\varphi)$, the labels associated with the prescribed
root $\zeta_0$ are precisely $(\lambda,\beta)$.  We call the
other two roots, counted with multiplicity, the rival roots.
Panels~(d)--(f) of Fig.~\ref{fig:construction-flow} show the root labels
along the magnetic axis, $\mathcal L_{\mathrm{core}}$, and
$\mathcal L_{\mathrm{out}}$, respectively.  In each panel, the prescribed
root gives one fixed label pair inside $\mathcal E$, while the two rival
roots trace phase-colored curves outside $\mathcal E$ as $\varphi$
traverses $S^1$.

Using interval arithmetic~\cite{moorekearfottcloud2009}, we establish, for every
$(\lambda,\beta,\varphi)\in\overline{\mathcal E}\times S^1$,
\begin{equation}
 Q\bigl(\lambda_\eta,
 \beta_\eta(\X(\lambda,\beta,\varphi))\bigr)\ge2
 \label{eq:main-rival-exclusion}
\end{equation}
for each rival root $\eta$, together with
\begin{equation}
 \left|D\bigl(\zeta_0;\X(\lambda,\beta,\varphi)\bigr)\right|^2\ge\frac1{100}.
 \label{eq:main-root-simple}
\end{equation}
A Python program verifies Eqs.~\eqref{eq:main-rival-exclusion}
and~\eqref{eq:main-root-simple} throughout each box of a finite cover of
$\overline{\mathcal E}\times S^1$.
Supplemental Section~\ref{sec:supp-interval-certificate} describes the
algorithm; the code is in the accompanying
repository~\cite{cubicRootCertificate}.

At each point $\bm x=\X(\lambda,\beta,\varphi)$ with
$(\lambda,\beta,\varphi)\in\overline{\mathcal E}\times S^1$,
Eqs.~\eqref{eq:main-rival-exclusion} and~\eqref{eq:main-root-simple}
ensure that the prescribed root $\zeta_0$ is simple and is the unique root
whose associated labels lie in $\overline{\mathcal E}$.  This root fixes
$\lambda=|\zeta_0|^2$, $\varphi=\arg\zeta_0$ modulo $2\pi$, and
$\beta=\beta_{\zeta_0}(\bm x)$.  Thus each image point $\bm x$ determines
its parameter triple $(\lambda,\beta,\varphi)$ uniquely, proving that
$\X$ is injective on $\overline{\mathcal E}\times S^1$.

We can also determine whether a spatial point $\bm x$ belongs to the
constructed domain directly from the roots of the cubic polynomial at $\bm x$.  For any nonzero root
$\eta=\sqrt{\lambda_\eta}\,e^{i\varphi_\eta}$ at $\bm x$,
$G(\eta;\bm x)=0$ and the definition of $\beta_\eta(\bm x)$ in
Eq.~\eqref{eq:R} give the same nonsingular linear system used to define
$\X$, with $\zeta_0$ replaced by $\eta$.  Its unique solution is therefore
\begin{equation}
 \bm x=\X\bigl(\lambda_\eta,\beta_\eta(\bm x),\varphi_\eta\bigr).
 \label{eq:main-coordinate-recovery}
\end{equation}
A point $\bm x$ therefore lies in $\X(\overline{\mathcal E}\times S^1)$
exactly when some root at $\bm x$ has associated labels in
$\overline{\mathcal E}$.

On $\mathcal P=\overline{\mathcal E}\times S^1$, the Jacobian of the
coordinate map satisfies
\begin{equation}
 \det[\partial_\lambda\X,\partial_\beta\X,\partial_\varphi\X]
 =\frac{\left|D\bigl(\zeta_0;\X(\lambda,\beta,\varphi)\bigr)\right|^2}
 {2(1+\lambda)^2}>0.
 \label{eq:main-coordinate-jacobian}
\end{equation}
The strict inequality follows from Eq.~\eqref{eq:main-root-simple}.
For the equality, set $\sigma=\sqrt\lambda$.  Since
$\partial_\beta\X=\nn$, direct calculation gives
\begin{equation}
 \sigma\det[\partial_\sigma\X,\partial_\beta\X,\partial_\varphi\X]
 =|\partial_\varphi\X|^2=\frac{\lambda|D|^2}{(1+\lambda)^2},
 \label{eq:main-jacobian-identities}
\end{equation}
and $\partial_\sigma\X=2\sigma\,\partial_\lambda\X$ then gives
Eq.~\eqref{eq:main-coordinate-jacobian}.
The formula for $\X$ is real analytic on
$(0,\infty)\times\RR\times S^1$.
Equation~\eqref{eq:main-coordinate-jacobian} and global injectivity on
$\mathcal P$ make
\begin{equation}
 \X:\mathcal E\times S^1
 \longrightarrow\X(\mathcal E\times S^1)
 \label{eq:main-global-diffeomorphism}
\end{equation}
a global real-analytic diffeomorphism.  The physical domain is
\begin{equation}
 \Omega=\X(\mathcal E\times S^1).
 \label{eq:main-domain}
\end{equation}
The restriction $\X|_{\mathcal P}$ is an embedding.  Since $\mathcal P$ is
compact and $\mathcal E\times S^1$ is dense in it,
$\overline\Omega=\X(\mathcal P)$ is an embedded compact solid torus.

The inverse in Eq.~\eqref{eq:main-global-diffeomorphism},
$\bm x\mapsto(\lambda(\bm x),\beta(\bm x),\varphi(\bm x))$, defines on
$\Omega$ the global nonzero real-analytic root
$\zeta(\bm x)=\sqrt{\lambda(\bm x)}\,e^{i\varphi(\bm x)}$ used in
Eq.~\eqref{eq:B}, with $\lambda(\bm x)=|\zeta(\bm x)|^2$ and
$\beta(\bm x)=\beta_\zeta(\bm x)$ as defined above.  On $\Omega$, the
flux surface label is
\begin{equation}
 \psi(\bm x)=Q(\lambda(\bm x),\beta(\bm x)).
 \label{eq:psi}
\end{equation}
The analyticity of $\X$, its injectivity on the compact set $\mathcal P$,
and its nonzero Jacobian there imply that its inverse extends real
analytically to an open neighborhood of $\overline\Omega$.  Hence
$\zeta$, $\lambda$, $\beta$, $\psi$, and $\B$ extend real analytically
across $\partial\Omega$.
Cardano's formulas give the three roots of Eq.~\eqref{eq:cubic}.  By the
root separation bound in Eq.~\eqref{eq:main-rival-exclusion}, the condition
$(\lambda_\eta,\beta_\eta(\bm x))\in\mathcal E$ selects exactly one throughout
$\Omega$.
Supplemental Section~\ref{subsec:cardano-root-formula} uses this criterion
to give finite closed-form Cartesian expressions for $\B$, $\psi$, and
$\Omega$.

\subsection{First integrals and flux function}
\label{subsec:main-vacuum}
\label{subsec:main-flux-regularity}

Here $D=\partial_\zeta G$ is evaluated at $(\zeta(\bm x),\bm x)$.
Equation~\eqref{eq:B} gives the
squared field strength
\begin{equation}
 |\B|^2=\frac{(1+\lambda)^2}{\lambda|D|^2}>0.
 \label{eq:main-nonzero}
\end{equation}
The strict inequality shows that $\B$ does not vanish anywhere in $\Omega$.

Differentiating Eq.~\eqref{eq:main-forward} and using Eq.~\eqref{eq:B} gives
\begin{equation}
 \B\bigl(\X(\lambda,\beta,\varphi)\bigr)
 =\frac{\partial_\varphi\X(\lambda,\beta,\varphi)}
 {|\partial_\varphi\X(\lambda,\beta,\varphi)|^2}.
 \label{eq:main-forward-identities}
\end{equation}
Applying the chain rule to the inverse coordinate map and substituting
Eqs.~\eqref{eq:main-coordinate-jacobian},
\eqref{eq:main-jacobian-identities}, and
\eqref{eq:main-forward-identities}, we obtain
\begin{equation}
 \grad\lambda\times\grad\beta
 =\frac{\partial_\varphi\X}
 {\det[\partial_\lambda\X,\partial_\beta\X,\partial_\varphi\X]}
 =2\lambda\B.
 \label{eq:main-clebsch-proof}
\end{equation}
Since $\lambda=|\zeta|^2>0$, the magnetic field admits the global
Clebsch representation~\cite{helander2014}
\begin{equation}
 \B=\grad\log|\zeta|\times\grad\beta
 =\frac{\grad\lambda\times\grad\beta}{2\lambda}.
 \label{eq:main-clebsch}
\end{equation}
The Clebsch potentials $\log|\zeta|=\tfrac12\log\lambda$ and $\beta$ are
single-valued and real analytic throughout $\Omega$, including the magnetic
axis, since $\lambda\geq7/60>0$ and the inverse coordinate map is real
analytic.
Equation~\eqref{eq:main-clebsch} implies
$\B\mathbin{\cdot}\grad\lambda=\B\mathbin{\cdot}\grad\beta=0$, so $\lambda$
and $\beta$ are constant along magnetic field lines.
Since $\grad\lambda\times\grad\beta=2\lambda\B\ne\bm0$, their gradients
are linearly independent.  Together with the global coordinates of
Eq.~\eqref{eq:main-global-diffeomorphism} and
Eq.~\eqref{eq:main-forward-identities}, these identities show that each pair
$(\lambda,\beta)\in\mathcal E$ labels a unique closed magnetic field line.
Since $\psi$ depends on $\bm x$ only through $\lambda$ and $\beta$,
$\B\mathbin{\cdot}\grad\psi=0$, so $\psi$ is a flux function
and its level sets are flux surfaces.
Differentiating Eq.~\eqref{eq:psi} and using the independence of
$\grad\lambda$ and $\grad\beta$ shows that
$\grad\psi=0$ exactly on the closed curve where
$(\lambda,\beta)=(\lambda_0,\beta_0)$, which is the magnetic axis $\K$ given by
Eq.~\eqref{eq:main-axis}, and that $\grad\psi$ is nonzero on every level
$0<\psi<1$.

\subsection{Magnetic axis and nested toroidal flux surfaces}
\label{sec:main-geometry}

The center of $\mathcal E$ in Eq.~\eqref{eq:main-parameter-center} gives the
embedded closed field line
\begin{equation}
 \K(\varphi)=\X(\lambda_0,\beta_0,\varphi).
 \label{eq:main-axis}
\end{equation}
On this curve $\psi=0$, and the Hessian of $\psi$ has quadratic form
$1800(\delta\bm x\mathbin{\cdot}\grad\lambda)^2
+200(\delta\bm x\mathbin{\cdot}\grad\beta)^2$ on a displacement $\delta\bm x$.  By the
independence of $\grad\lambda$ and $\grad\beta$, this expression vanishes
exactly when $\delta\bm x$ is tangent to $\K$ and is positive for every other
displacement.  Thus $\K$ is a regular magnetic axis.

For $0<h<1$, parametrizing the ellipse $Q=h$ by $\alpha$ gives
\begin{equation}
 (\alpha,\varphi)\longmapsto
 \X\!\left(\frac3{20}+\frac{\sqrt h}{30}\cos\alpha,
             \frac1{50}+\frac{\sqrt h}{10}\sin\alpha,\varphi\right).
 \label{eq:main-surfaces}
\end{equation}
The image is an embedded compact connected toroidal flux surface, foliated
by the closed field lines at fixed $\alpha$.  The elliptic regions
$Q\le h$ give the nested solid tori
$\X(\{Q\le h\}\times S^1)$, which fill $\Omega$ and contract
to $\K$ as $h$ goes to zero.  The level $Q=1$ maps to
$\partial\Omega$; there
$\grad\psi$ is normal and nonzero, while
$\B\mathbin{\cdot}\grad\psi=0$, so $\B$ is tangent to the boundary.

\begin{figure*}[t]
\centering
\includegraphics[width=\textwidth]{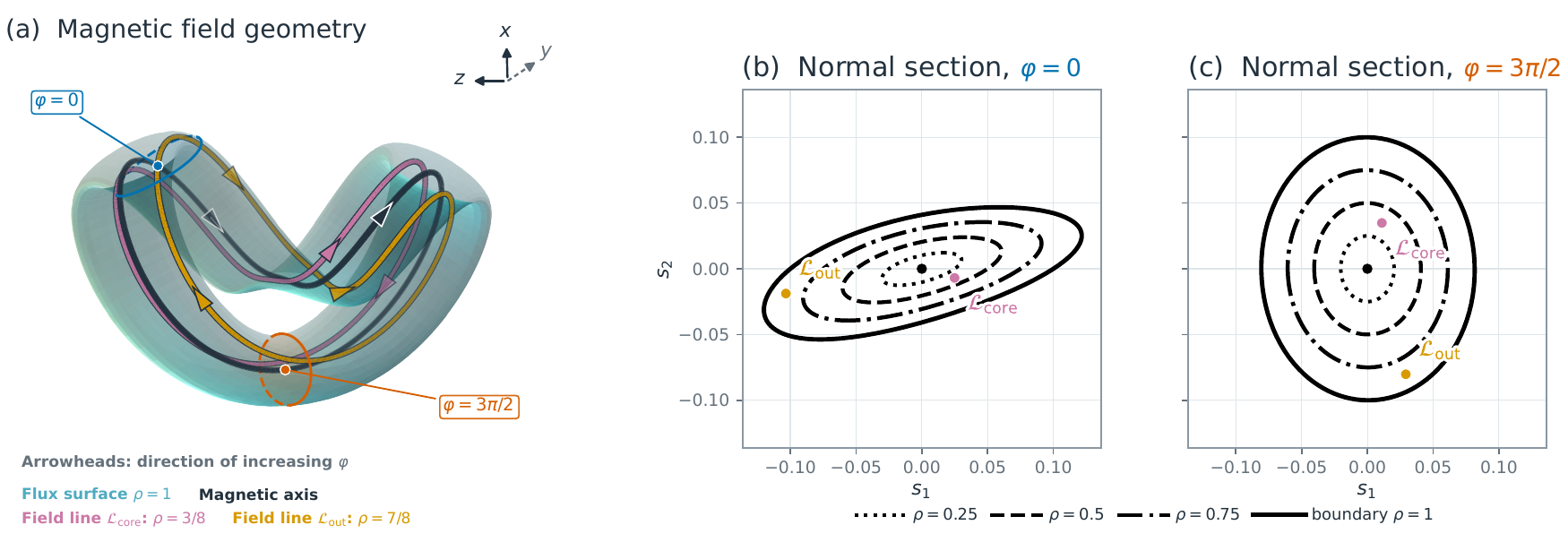}
\caption{Geometry and normal-plane sections of the nested flux surfaces in
$\Omega$.
The nested flux surfaces are labeled by the radial surface parameter
$\rho=\sqrt\psi$, with $\rho=\sqrt h$ in Eq.~\eqref{eq:main-surfaces}.
(a) The boundary flux surface $\rho=1$, magnetic axis $\K$, and the
representative field lines $\mathcal L_{\mathrm{core}}$ and
$\mathcal L_{\mathrm{out}}$, with $\rho=3/8$ and $7/8$, respectively.
Arrowheads point along $\partial_\varphi\X$, the direction of increasing
$\varphi$ and, by Eq.~\eqref{eq:main-forward-identities}, of $\B$.  The
blue and orange footprints mark the normal planes at
$\varphi_0=0$ and $\varphi_0=3\pi/2$.
(b), (c) Intersections of the surfaces $\rho=0.25$, $0.5$, $0.75$, and $1$
with the normal plane through $\K(\varphi_0)$, at $\varphi_0=0$ and $3\pi/2$,
respectively.  Here $(s_1,s_2)$ are the components of $\bm x-\K(\varphi_0)$ along
$\bm e_1(\varphi_0)$ and $\bm e_2(\varphi_0)$ of Eq.~\eqref{eq:main-geometric-frame}
for each intersection point $\bm x$, and the black dot marks $\K(\varphi_0)$.
The nested sections are nearly similar ellipses, more elongated at
$\varphi_0=0$ than at $\varphi_0=3\pi/2$.
The phase $\varphi$ varies along each section curve.}
\label{fig:geometry-sections}
\end{figure*}

Figure~\ref{fig:geometry-sections}(a) displays the boundary $\partial\Omega$,
the magnetic axis $\K$, and representative field lines.
Figures~\ref{fig:geometry-sections}(b) and (c) show sections in the normal
planes through $\K(\varphi_0)$, with $\varphi_0=0$ and $3\pi/2$,
respectively.
For these sections, use the orthonormal frame adapted to the
axis tangent $\K'(\varphi)$:
\begin{equation}
 \widehat{\bm T}=\frac{\K'}{|\K'|},\qquad
 \bm e_1=\frac{\bm e_y\times\widehat{\bm T}}
 {|\bm e_y\times\widehat{\bm T}|},\qquad
 \bm e_2=\widehat{\bm T}\times\bm e_1,
 \label{eq:main-geometric-frame}
\end{equation}
where $\bm e_y=(0,1,0)$.  The axis has a simple closed projection onto the
$xz$-plane with a nonzero tangent, as checked in Supplemental
Section~\ref{sec:supp-axis-projection}.  Hence
$|\bm e_y\times\widehat{\bm T}|>0$, and normalization gives a smooth
basis defined everywhere along the axis.

\subsection{Rotational transform \texorpdfstring{$\iota=-1$}{iota = -1}}
\label{sec:main-transform}

The rotational transform counts signed poloidal turns per toroidal transit.
Each field line in Eq.~\eqref{eq:main-surfaces} has constant $\alpha$ and
closes as $\varphi$ increases by $2\pi$, completing one toroidal transit
in the direction of $\B$ by Eq.~\eqref{eq:main-forward-identities}.
We compute the winding of one field line near the magnetic axis and then extend
its integer value by continuity.

The basis in Eq.~\eqref{eq:main-geometric-frame} fixes the reference from
which the geometric poloidal angle is measured.  The magnetic axis is the
graph of its $y$ coordinate over its simple $xz$-projection.  Continuously
scaling the axis's $y$ coordinate to zero flattens the axis into the
$xz$-plane.  After flattening, $\bm e_1$ lies in the $xz$-plane and is
perpendicular to the flattened magnetic axis's tangent.  A sufficiently
small displacement along $\bm e_1$ traces a closed curve in the $xz$-plane
alongside the flattened axis, making zero poloidal turns about the
flattened axis.  Restoring the axis's $y$ coordinate preserves the zero
poloidal turn count.  The basis therefore introduces no additional net
poloidal winding.  The positive orientation is
$\bm e_1\times\bm e_2=\widehat{\bm T}$.

For sufficiently small $\epsilon>0$, consider the field line
$\bm x_\epsilon(\varphi)=\X(\lambda_0,\beta_0+\epsilon,\varphi)$.
Equations~\eqref{eq:main-frame-data} and~\eqref{eq:main-forward} give its
exact displacement from the magnetic axis:
\begin{equation}
 \begin{aligned}
 \bm x_\epsilon-\K&=\epsilon\nn_0(\varphi),\\
 \nn_0(\varphi)&=\frac{(4\sqrt{15}\cos\varphi,\,
                         4\sqrt{15}\sin\varphi,\,-17)}{23}.
 \end{aligned}
 \label{eq:main-calibration-line}
\end{equation}
By Eq.~\eqref{eq:main-coordinate-normal}, $\nn_0$ is a unit vector in the
normal plane to the axis.  Encode the displacement's components along
$\bm e_1$ and $\bm e_2$ as the real and imaginary parts of
\begin{equation}
 \begin{aligned}
 \chi_\epsilon
 &=(\bm x_\epsilon-\K)\mathbin{\cdot}\bm e_1
   +i(\bm x_\epsilon-\K)\mathbin{\cdot}\bm e_2,\\
 \theta_{\mathrm g}&=\arg\chi_\epsilon.
 \end{aligned}
 \label{eq:main-geometric-poloidal-angle}
\end{equation}
Its modulus is $\epsilon$, so its angle can be followed continuously.
Substitution of Eqs.~\eqref{eq:main-geometric-frame}
and~\eqref{eq:main-calibration-line}, with $\K'$ obtained from
Eq.~\eqref{eq:main-forward}, gives
\begin{equation}
 \begin{aligned}
 \ImPart\chi_\epsilon
 &=\frac{4\epsilon\sqrt{15}\,|\K'|}
 {23\,|\bm e_y\times\K'|}\sin\varphi,\\
 \chi_\epsilon(0)&=\chi_\epsilon(2\pi)=-\epsilon,\qquad
 \chi_\epsilon(\pi)=\epsilon.
 \end{aligned}
 \label{eq:main-winding-signs}
\end{equation}
For $0<\varphi<\pi$, the displacement lies in the upper half-plane and runs from the negative to the positive real axis.
For $\pi<\varphi<2\pi$, it returns through the lower half-plane.
The continuously followed angle therefore goes from $\pi$ to $0$ to
$-\pi$, giving
\begin{equation}
 \Delta\theta_{\mathrm g}=-2\pi,\qquad \Delta\varphi=2\pi.
 \label{eq:main-geometric-angle-change}
\end{equation}
This field line consequently has $\iota=-1$.

Any other off-axis field line can be reached from the reference line by
continuously varying $(\lambda,\beta)$ within $\mathcal E$ while avoiding
the magnetic axis labels $(\lambda_0,\beta_0)$.  Each field line continues
to close after one toroidal transit, so its number of poloidal turns
around the magnetic axis is an integer.  The integer cannot change during
the continuous deformation unless the field line crosses the magnetic
axis.  The embedding in Eq.~\eqref{eq:main-global-diffeomorphism} keeps
every off-axis field line disjoint from the magnetic axis.  Hence
\begin{equation}
 \iota(h)=-1,\qquad 0<h<1.
 \label{eq:main-iota}
\end{equation}

\subsection{Absence of continuous Euclidean symmetry}
\label{subsec:main-symmetry}

An infinitesimal generator of a Euclidean motion has the form
$\boldsymbol{\xi}(\bm x)=\bm a+\bm\omega\times\bm x$, where
$\bm a,\bm\omega\in\RR^3$ are constant real vectors generating translations
and rotations, respectively.  Invariance of $\B$ under this motion requires
\begin{equation}
 (\boldsymbol{\xi}\mathbin{\cdot}\grad)\B-\bm\omega\times\B=0
 \qquad\text{for all }\bm x\in\Omega.
 \label{eq:main-infinitesimal-generator}
\end{equation}
Evaluating Eq.~\eqref{eq:main-infinitesimal-generator} at two points in $\Omega$
gives six homogeneous linear equations for the six components of
$(\bm a,\bm\omega)$.  Take
$\bm p_1=\X(25/169,1/50,0)$ and
$\bm p_2=\X(25/169,3/100,0)$.  At both points $\zeta=5/13$ and $Q<1$,
so both lie in $\Omega$.  Direct substitution of the field expressions
using exact rational arithmetic
shows that the determinant of the resulting $6\times6$ coefficient matrix
is nonzero, with approximate value $3.2\times10^{-3}$, so
$\bm a=\bm\omega=\bm0$.  The left-hand side of
Eq.~\eqref{eq:main-infinitesimal-generator} is real analytic throughout the
connected domain $\Omega$.  If a nontrivial Euclidean symmetry existed on any open subset of
$\Omega$, this analytic expression would vanish there and hence, by the
identity theorem, throughout $\Omega$, contradicting the two-point
calculation.  Thus the field has neither local nor global continuous Euclidean
symmetry.

\section{Discussion}\label{sec:conclusion}

We have constructed an exact vacuum magnetic field without continuous
Euclidean symmetry that fills a bounded solid torus with nested toroidal
flux surfaces and nonzero rotational transform.  The
field is nowhere vanishing, is analytic through a regular magnetic axis,
and has $\iota=-1$ on every flux surface.

The closed-form field and flux surfaces provide an exact benchmark for
three-dimensional equilibrium codes.  VMEC~\cite{hirshmanwhitson1983} and
DESC~\cite{dudtkolemen2020} require one closed poloidal contour at each
geometric toroidal angle in their cylindrical $R,Z$ representation.
In a suitably chosen cylindrical frame, the smaller domain
$\psi<1/100$ [Fig.~\ref{fig:supp-family}(d)] is a candidate for a
fixed-boundary VMEC calculation.  The geometries obtained here may
also provide benchmarks for GVEC, whose general coordinate
frame~\cite{hindenlang2026gvec} accommodates knotted magnetic axes
[Fig.~\ref{fig:supp-family}(c)].

Numerical exploration of other Weierstrass
polynomials~\cite{bairdwood1988,bairdwood1997} and field line label domains
suggests additional configurations, including one with a knotted magnetic
axis~\cite{hudson2014} and a rotational
transform consistent with $\iota=+3$.
The field is curl free and divergence free wherever the selected root is
nonzero and simple, by Eq.~\eqref{eq:main-harmonic-chain}; an embedded solid
torus additionally requires global injectivity of the coordinate map.
Global injectivity is certified for the principal cubic configuration by
the computer-assisted proof in Supplemental
Section~\ref{sec:supp-interval-certificate}.

Every off-axis field line in this construction closes after one toroidal transit
of $\varphi=\arg\zeta$, so the rotational transform is an integer and is
constant across the nested surfaces.
Closed-form finite-pressure equilibria already realize both
a constant rotational transform of $\iota=+2$ and a sheared transform
profile~\cite{landreman}.  Whether a vacuum field without continuous
Euclidean symmetry can fill a solid torus with nested toroidal flux surfaces
and a sheared transform profile is open.

\section*{Use of large language models and agentic tools}

These solutions were discovered by a team of AI agents built on
Claude Sonnet 5.5, Claude Opus 5.5, GPT-5.6 Sol, and GPT-6 Astra and directed
by the author.  GPT-6 Astra proposed the connection to
Ref.~\cite{bairdwood1997}.  The agents also assisted with code and with
preparing the manuscript.  All equations were confirmed by the author.

\section*{Acknowledgments}

The author thanks Mark Cianciosa for discussions on boundary shapes suitable for VMEC
and on the low-order rational $\iota=-1$.
This manuscript has been authored by UT-Battelle, LLC, under contract DE-AC05-00OR22725 with the US Department of Energy (DOE). 
The US government retains and the publisher, by accepting the work for publication, 
acknowledges that the US government retains a non-exclusive, paid-up, irrevocable, world-wide license to publish or reproduce the submitted manuscript version of this work, or allow others to do so, for US government purposes. DOE will provide public access to these results of federally sponsored research in accordance with the DOE Public Access Plan (http://energy.gov/downloads/doe-public-access-plan).

\bibliographystyle{apsrev4-2}
\bibliography{references}

@misc{landreman,
  title={Analytic toroidal {3D MHD} equilibria and steady {Euler} flows with invariant surfaces},
  author={Landreman, Matt},
  eprint={2609.26742},
  archivePrefix={arXiv},
  primaryClass={physics.plasm-ph},
  year={2026}
}

@misc{issan2026,
  title={Analytic non-axisymmetric {3D MHD} equilibria without stellarator symmetry},
  author={Issan, Opal and Citrin, Jonathan and Landreman, Matt and Tracey, Brendan},
  eprint={2610.07304},
  archivePrefix={arXiv},
  primaryClass={physics.plasm-ph},
  year={2026}
}

@article{weitzner,
  title={Exact non-symmetric closed line vacuum magnetic fields in a topological torus},
  author={Weitzner, Harold and Sengupta, Wrick},
  journal={Physics of Plasmas},
  volume={27},
  number={2},
  pages={022509},
  year={2020},
  publisher={AIP Publishing}
}

@article{helander2014,
  title={Theory of plasma confinement in non-axisymmetric magnetic fields},
  author={Helander, Per},
  journal={Reports on Progress in Physics},
  volume={77},
  number={8},
  pages={087001},
  year={2014},
  publisher={IOP Publishing}
}

@article{cary1982,
  title={Vacuum magnetic fields with dense flux surfaces},
  author={Cary, John R},
  journal={Physical Review Letters},
  volume={49},
  number={4},
  pages={276},
  year={1982},
  publisher={APS}
}

@article{cary1984,
  title={Construction of three-dimensional vacuum magnetic fields with dense nested flux surfaces},
  author={Cary, John R},
  journal={Physics of Fluids},
  volume={27},
  number={1},
  pages={119--128},
  year={1984},
  publisher={American Institute of Physics}
}

@article{hansoncary1984,
    author = {Hanson, James D. and Cary, John R.},
    title = {Elimination of stochasticity in stellarators},
    journal = {Physics of Fluids},
    volume = {27},
    number = {4},
    pages = {767-769},
    year = {1984},
    month = {04},
    issn = {0031-9171},
}

@article{giuliani2023,
  title={Direct stellarator coil optimization for nested magnetic surfaces with precise quasi-symmetry},
  author={Giuliani, Andrew and Wechsung, Florian and Cerfon, Antoine and Landreman, Matt and Stadler, Georg},
  journal={Physics of Plasmas},
  volume={30},
  number={4},
  pages={042511},
  year={2023},
  publisher={AIP Publishing}
}

@article{lortz1970,
  author = {Lortz, Dietrich},
  title = {{\"U}ber die Existenz toroidaler magnetohydrostatischer Gleichgewichte ohne Rotationstransformation},
  journal = {Zeitschrift f{\"u}r Angewandte Mathematik und Physik},
  volume = {21},
  number = {2},
  pages = {196--211},
  year = {1970},
}

@article{satoyamada2023,
  title={Nested invariant tori foliating a vector field and its curl: Toward MHD equilibria and steady Euler flows in toroidal domains without continuous Euclidean isometries},
  author={Sato, Naoki and Yamada, Michio},
  journal={Journal of Mathematical Physics},
  volume={64},
  number={8},
  pages={081505},
  year={2023},
  publisher={AIP Publishing}
}

@article{bairdwood1988,
  title={Bernstein theorems for harmonic morphisms from $\mathbb{R}^3$ and $S^3$},
  author={Baird, P and Wood, John C},
  journal={Mathematische Annalen},
  volume={280},
  pages={579--603},
  year={1988}
}

@article{bairdwood1997,
  title={Weierstrass representations for harmonic morphisms on Euclidean spaces and spheres},
  author={Baird, P and Wood, John C},
  journal={Mathematica Scandinavica},
  volume={81},
  pages={283--300},
  year={1997},
  publisher={JSTOR}
}

@article{grad1967,
  title={Toroidal containment of a plasma},
  author={Grad, Harold},
  journal={Physics of Fluids},
  volume={10},
  number={1},
  pages={137--154},
  year={1967},
  publisher={AIP Publishing}
}

@article{grad1985,
  author = {Grad, Harold},
  title = {Theory and applications of the nonexistence of simple toroidal plasma equilibrium},
  journal = {International Journal of Fusion Energy},
  volume = {3},
  number = {2},
  pages = {33--46},
  year = {1985},
}

@article{constantin2021,
  title={Flexibility and rigidity in steady fluid motion},
  author={Constantin, Peter and Drivas, Theodore D and Ginsberg, Daniel},
  journal={Communications in Mathematical Physics},
  volume={385},
  number={1},
  pages={521--563},
  year={2021},
  publisher={Springer}
}

@misc{gomezserrano2026,
  title={Counterexamples to {Grad}'s conjecture},
  author={G{\'o}mez-Serrano, Javier and Liehr, Lukas and Taylor, Mitchell A},
  eprint={2609.24739},
  archivePrefix={arXiv},
  primaryClass={math.AP},
  year={2026}
}

@article{hirshmanwhitson1983,
  author = {Hirshman, S. P. and Whitson, J. C.},
  title = {Steepest-descent moment method for three-dimensional magnetohydrodynamic equilibria},
  journal = {Physics of Fluids},
  volume = {26},
  number = {12},
  pages = {3553},
  year = {1983}
}

@article{dudtkolemen2020,
  title={{DESC}: a stellarator equilibrium solver},
  author={Dudt, Daniel W and Kolemen, Egemen},
  journal={Physics of Plasmas},
  volume={27},
  number={10},
  pages={102513},
  year={2020},
  publisher={AIP Publishing}
}

@article{hindenlang2026gvec,
  title = {{GVEC}: A flexible {3D MHD} equilibrium solver},
  author={Hindenlang, Florian and Maj, Omar and Babin, Robert and K{\"o}berl, Robert and Muir, Dean and Ribeiro, Tiago Tamissa and Rampp, Markus and Sonnendr{\"u}cker, Eric},
  journal={Journal of Open Source Software},
  volume={11},
  number={120},
  pages={9670},
  year={2026}
}

@book{moorekearfottcloud2009,
  author = {Moore, Ramon E. and Kearfott, R. Baker and Cloud, Michael J.},
  title = {Introduction to Interval Analysis},
  publisher = {Society for Industrial and Applied Mathematics},
  address = {Philadelphia, PA},
  year = {2009},
}

@misc{cubicRootCertificate,
  author = {L\'opez, O. E.},
  title = {Certified root separation for a cubic polynomial},
  howpublished = {\url{https://github.com/omlopez/cubicRootCertificate}},
  year = {2026}
}

@article{senguptaweitzner2019,
  title={Low-shear three-dimensional equilibria and vacuum magnetic fields with flux surfaces},
  author={Sengupta, Wrick and Weitzner, Harold},
  journal={Journal of Plasma Physics},
  volume={85},
  number={2},
  pages={905850209},
  year={2019},
  publisher={Cambridge University Press}
}

@article{hudson2014,
  title={A new class of magnetic confinement device in the shape of a knot},
  author={Hudson, S. R. and Startsev, E. and Feibush, E.},
  journal={Physics of Plasmas},
  volume={21},
  number={1},
  pages={010705},
  year={2014},
  publisher={AIP Publishing}
}

@article{brunolaurence1996,
  title={Existence of three-dimensional toroidal {MHD}  equilibria with nonconstant pressure},
  author={Bruno, Oscar P and Laurence, Peter},
  journal={Communications on Pure and Applied Mathematics},
  volume={49},
  number={7},
  pages={717--764},
  year={1996},
  publisher={Wiley Online Library}
}

@article{enciso2025,
  title={{MHD} equilibria with nonconstant pressure in nondegenerate toroidal domains},
  author={Enciso, Alberto and Luque, Alejandro and Peralta-Salas, Daniel},
  journal={Journal of the European Mathematical Society},
  volume={27},
  number={6},
  pages={2251--2291},
  year={2025}
}

@article{boozer2019curlfree,
  title={Curl-free magnetic fields for stellarator optimization},
  author={Boozer, Allen H.},
  journal={Physics of Plasmas},
  volume={26},
  number={10},
  pages={102504},
  year={2019},
}

\onecolumngrid
\clearpage
\section*{Supplemental Material}
\setcounter{section}{0}
\setcounter{subsection}{0}
\setcounter{equation}{0}
\setcounter{figure}{0}
\setcounter{table}{0}
\renewcommand{\thesection}{S\arabic{section}}
\renewcommand{\thesubsection}{S\arabic{section}.\arabic{subsection}}
\renewcommand{\theequation}{S\arabic{equation}}
\renewcommand{\thefigure}{S\arabic{figure}}
\renewcommand{\thetable}{S\arabic{table}}
\renewcommand{\theHsection}{supp.\arabic{section}}
\renewcommand{\theHsubsection}{supp.\arabic{section}.\arabic{subsection}}
\renewcommand{\theHequation}{supp.\arabic{equation}}
\renewcommand{\theHfigure}{supp.\arabic{figure}}
\renewcommand{\theHtable}{supp.\arabic{table}}
\makeatletter
\renewcommand{\p@subsection}{}
\makeatother

\section{Certified global root separation}
\label{sec:supp-interval-certificate}

We certify Eqs.~\eqref{eq:main-rival-exclusion}
and~\eqref{eq:main-root-simple} on the closed parameter set
\begin{equation}
 \widetilde{\mathcal P}=\{(u,v,\beta):Q(u^2+v^2,\beta)\le1\},\qquad
 \zeta_0=u+iv,\qquad \lambda=u^2+v^2.
 \label{eq:supp-certificate-domain}
\end{equation}
On $\widetilde{\mathcal P}$, $7/60\le\lambda\le11/60$.  The lower bound keeps $\zeta_0$
away from zero, so $(u,v,\beta)\in\widetilde{\mathcal P}$ corresponds one-to-one to
$(\lambda,\beta,\varphi)\in\mathcal P$ through
$u+iv=\sqrt\lambda\,e^{i\varphi}$.  The upper bound gives $|u|,|v|<1/2$, so the initial box
$[-1/2,1/2]^2\times[-2/25,3/25]$, with its rational endpoints rounded
outward, contains all of $\widetilde{\mathcal P}$.

All computations use interval arithmetic: each real quantity is replaced
by an interval that contains its value at every point of the current box,
and each complex quantity by a pair of such intervals.  The accompanying
Python program~\cite{cubicRootCertificate} stores endpoints exactly as
integer multiples of $2^{-80}$, rounding lower endpoints down and upper
endpoints up.

For each box, we evaluate Eq.~\eqref{eq:main-forward} in its rational
$(u,v,\beta)$ form to enclose $\bm x=(x,y,z)$, and enclose
$|D(\zeta_0;\bm x)|^2$.  Since $G(\zeta_0;\bm x)=0$,
division by $Z-\zeta_0$ leaves a quadratic whose roots are precisely the
two rivals.  With $w=x+iy$, we have
\begin{equation}
 \frac{G(Z;\bm x)}{2i}
 =(Z-\zeta_0)(Z^2+LZ+C_q),\qquad
 L=\zeta_0+\frac{\bar w+2}{2i},\qquad C_q=\zeta_0L+iz.
 \label{eq:supp-certificate-factor}
\end{equation}
The rival roots are therefore
\begin{equation}
 \eta_\pm=\frac{-L\pm\sqrt{\Delta_q}}2,\qquad
 \Delta_q=L^2-4C_q.
 \label{eq:supp-certificate-rivals}
\end{equation}
Interval arithmetic encloses both rival roots, including on boxes where
$\Delta_q$ may vanish.  For each enclosure, we recover
$\lambda_\eta=|\eta|^2$ and $\beta_\eta(\bm x)$ from Eq.~\eqref{eq:R}
and bound $Q(\lambda_\eta,\beta_\eta(\bm x))$.

A box is discarded only if the interval lower bound for
$Q(u^2+v^2,\beta)$ exceeds $1$.  Every other box is certified only
if its lower bounds satisfy $|D(\zeta_0;\bm x)|^2\ge1/100$ and
$Q(\lambda_{\eta_\pm},\beta_{\eta_\pm}(\bm x))\ge2$ for both rival roots.
Otherwise, the box is bisected into two closed halves.  Bisection
preserves coverage, discarded boxes do not intersect $\widetilde{\mathcal P}$,
and each accepted bound holds throughout its box.  The computation completed
with every retained box certified and no unresolved boxes, establishing
Eqs.~\eqref{eq:main-rival-exclusion} and~\eqref{eq:main-root-simple}
throughout $\widetilde{\mathcal P}$.

\section{Finite Cartesian formula}
\label{subsec:cardano-root-formula}

Cardano's formula gives the three roots of Eq.~\eqref{eq:cubic} directly
from the Cartesian coordinates.  With $w=x+iy$, define
\begin{equation}
\begin{gathered}
 a_c=\frac{\bar w+2}{2i},\qquad p_c=iz-\frac{a_c^2}{3},\qquad
 q_c=\frac{2a_c^3}{27}-\frac{a_ciz}{3}+\frac{iw}{2},\\
 \delta_c=\frac{q_c^2}{4}+\frac{p_c^3}{27},\qquad
 c_\pm=-\frac{q_c}{2}\pm\sqrt{\delta_c}.
\end{gathered}
\label{eq:supp-cardano-data}
\end{equation}
Use the principal square root and choose $c$ as the larger-magnitude value
of $c_\pm$, taking $c_+$ in a tie.  Set $U=c^{1/3}$ using the principal
complex cube root.  The three roots are
\begin{equation}
 r_k=-\frac{a_c}{3}+\omega^kU-\frac{p_c}{3\omega^kU},
 \qquad k=0,1,2,\qquad \omega=\frac{-1+i\sqrt3}{2}.
 \label{eq:supp-cardano-roots}
\end{equation}

For each root $r_k=u_k+iv_k$, set $\lambda_k=|r_k|^2$ and evaluate $\beta_k$
from Eq.~\eqref{eq:R} using $(u_k,v_k,\lambda_k)$.  Each root thus has
associated labels $(\lambda_k,\beta_k)$ and a value $Q_k=Q(\lambda_k,\beta_k)$.
The condition $Q_k<1$ places these labels inside the ellipse $\mathcal E$.
The domain is therefore
\begin{equation}
 \Omega=\{\bm x:\min_{0\le k\le2}Q_k(\bm x)<1\},\qquad
 \overline\Omega=\{\bm x:\min_{0\le k\le2}Q_k(\bm x)\le1\}.
 \label{eq:supp-cartesian-domain}
\end{equation}
By the root separation bound in Eq.~\eqref{eq:main-rival-exclusion},
exactly one root at each $\bm x\in\Omega$ has $Q_k<1$, while both rival
roots have $Q_k>1$.  The smallest candidate value therefore belongs to the
selected root $\zeta(\bm x)$ and equals the flux surface label defined in
Eq.~\eqref{eq:psi}:
\begin{equation}
 \psi(\bm x)=Q(\lambda(\bm x),\beta(\bm x))
 =\min_{0\le k\le2}Q_k(\bm x),\qquad \bm x\in\Omega.
 \label{eq:supp-cartesian-flux}
\end{equation}
On the boundary, select the unique root with $Q_k\le1$.  Substitution of
the selected root in Eq.~\eqref{eq:B} gives the field.  This root is nonzero
and simple throughout $\overline\Omega$.  Radical branch choices can
permute the indices $k$, leaving the selected root and the smooth
flux surface label unchanged.

\section{Projection of the magnetic axis}
\label{sec:supp-axis-projection}

The $xz$-projection of the magnetic axis fixes the geometric reference
used to count poloidal turns in Sec.~\ref{sec:main-transform}.
To check that this projection is a simple closed curve with a nonzero
tangent, measure its polar angle about
$(x_c,z_c)=(-7/20,-17/1150)$ using
\begin{equation}
 W(\varphi)=K_x(\varphi)+\frac7{20}
            -i\left(K_z(\varphi)+\frac{17}{1150}\right).
 \label{eq:supp-axis-projection}
\end{equation}
Set $u=\sqrt{3/20}\cos\varphi$ and
$v=\sqrt{3/20}\sin\varphi$.  Substitution of the axis labels into
Eq.~\eqref{eq:main-forward} and differentiation with
$u'=-v$, $v'=u$, and $u^2+v^2=3/20$ give
\begin{equation}
 \begin{aligned}
 \ImPart W&=-\frac{9600-15120v}{12167}\,u,\\
 \ImPart(\overline W W')&>0\qquad(0\le\varphi\le2\pi),\\
 W(\pi/2)&=\frac{41}{460}+\frac{3\sqrt{15}}{115}>0,\\
 W(3\pi/2)&=\frac{41}{460}-\frac{3\sqrt{15}}{115}<0.
 \end{aligned}
 \label{eq:supp-axis-projection-signs}
\end{equation}
Here $\ImPart(\overline WW')=3[p(v)+10664640\,u^3]/(5\cdot23^6)$ with
$p(v)=1612800000v^5-2048000000v^4+113295200v^3+303327720v^2-47928800v+4099221$.
Sturm's theorem shows that $p$ and $p^2-10664640^2(3/20-v^2)^3$ have no
zeros for $|v|\le\sqrt{3/20}$, and both are positive at $v=0$, so
$p(v)>10664640|u|^3$.
Since $9600-15120v>0$, the only real-axis crossings occur at
$u=0$, namely $\varphi=\pi/2$ and $3\pi/2$.
The strict inequality in Eq.~\eqref{eq:supp-axis-projection-signs}
ensures $W\ne0$, a nonzero projected tangent, and
$d\arg W/d\varphi>0$.  The two crossings show that this angle increases
by exactly $2\pi$ in one period.  Thus each ray from $(x_c,z_c)$ meets
the projection once, establishing the required simple closed curve.

\section{Further configurations and polynomial families}
\label{sec:supp-polynomial-program}

\begin{figure}[!ht]
\centering
\includegraphics[width=\textwidth]{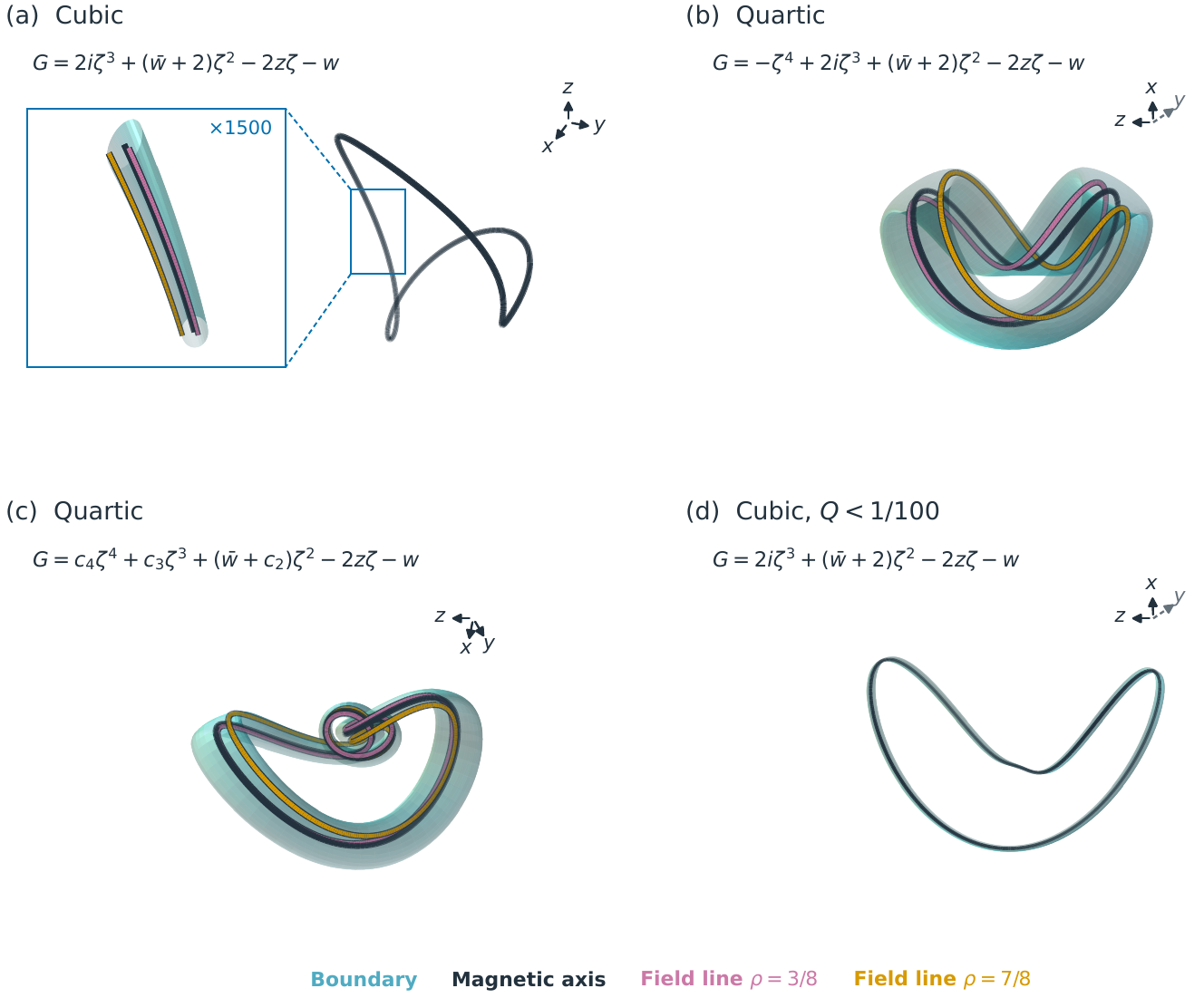}
\caption{Four vacuum configurations, with $\iota=-1$ in (d) and rotational
transforms consistent with $\iota=-1$ in (a) and (b) and $\iota=+3$ in (c).
(a) Configuration A, Eq.~\eqref{eq:supp-config-A}.  The domain is narrower
than the line width of the plotted magnetic axis; the zoom window shows the phase interval
$\varphi\in[4.4,4.7]$ with the same-phase displacement from the axis magnified
by $1500$.
(b) Configuration B, Eq.~\eqref{eq:supp-config-B}.
(c) Configuration C, Eq.~\eqref{eq:supp-config-C}.
In (a)--(c), $\rho=\sqrt{Q_A}$, $\sqrt{Q_B}$, and $\sqrt{Q_C}$,
respectively; the field lines, shown in (a) only within the zoom window,
have $\rho=3/8$ and $7/8$, at the same values
of $\alpha$ as $\mathcal L_{\mathrm{core}}$ and $\mathcal L_{\mathrm{out}}$.
(d) Configuration D, Eq.~\eqref{eq:supp-config-D}.}
\label{fig:supp-family}
\end{figure}

Figure~\ref{fig:supp-family} illustrates four further configurations
obtained by varying the $\bm x$-independent terms of Eq.~\eqref{eq:cubic} or the
elliptical domain $\mathcal E$ of field line labels in
Eq.~\eqref{eq:main-torus-parameter-ellipse}.  Each configuration uses
\begin{equation}
 G(\zeta;\bm x)=c_4\zeta^4+c_3\zeta^3+(\bar w+c_2)\zeta^2-2z\zeta-w ,
 \label{eq:supp-polynomial-family}
\end{equation}
where Eq.~\eqref{eq:cubic} has $(c_2,c_3,c_4)=(2,2i,0)$.  The terms
$c_2\zeta^2$, $c_3\zeta^3$, and $c_4\zeta^4$ do not involve $\bm x$, so $\grad_{\bm x}G=-\N$ is unchanged and
Eq.~\eqref{eq:main-harmonic-chain} holds with $D=\partial_\zeta G$
wherever the selected root is nonzero and simple.
For $\zeta_0=\sqrt\lambda e^{i\varphi}$ as in
Eq.~\eqref{eq:main-prescribed-root}, define
\begin{equation}
 \begin{aligned}
 g(\zeta)&=c_2\zeta^2+c_3\zeta^3+c_4\zeta^4,\\
 \kappa_g(\lambda,\varphi)
 &=\frac{2\lambda}{(1+\lambda)^2}\RePart\!\left(
 c_2\zeta_0+\frac{c_3}{2}\zeta_0^2+\frac{c_4}{3}\zeta_0^3\right).
 \end{aligned}
 \label{eq:supp-polynomial-coordinate-data}
\end{equation}
The field line label and coordinate map in Eqs.~\eqref{eq:R}
and~\eqref{eq:main-forward} apply with $H=g(\zeta_0)$ and
$\kappa=\kappa_g$.  The vectors $\bm A$, $\bm C$, and $\nn$
are defined as in Eq.~\eqref{eq:main-frame-data}.

Configuration A uses the cubic polynomial in Eq.~\eqref{eq:cubic} and the ellipse
\begin{equation}
 Q_A(\lambda,\beta)=30000^2\left[\left(\lambda-\frac23\right)^2+(\beta-1)^2\right]<1 .
 \label{eq:supp-config-A}
\end{equation}
Configuration B adds the quartic term $-\zeta^4$ to the cubic polynomial in Eq.~\eqref{eq:cubic} and uses the ellipse
\begin{equation}
 Q_B(\lambda,\beta)=900\left(\lambda-\frac3{20}\right)^2+64\beta^2<1 .
 \label{eq:supp-config-B}
\end{equation}
Configuration C has $c_2=-\frac{57}{50}+\frac{339}{500}i$,
$c_3=\frac{2083}{1000}+\frac{21}{250}i$, and
$c_4=-\frac{2621}{1000}-\frac{503}{250}i$, and uses
\begin{equation}
 Q_C(\lambda,\beta)=169\left(\lambda-\frac{129}{250}\right)^2
 +64\left(\beta+\frac{17}{1000}\right)^2<1 .
 \label{eq:supp-config-C}
\end{equation}
Numerical calculations indicate that its magnetic axis is knotted and its
rotational transform is $\iota=+3$, suggesting an example of a
knotatron~\cite{hudson2014}.

Configuration D uses the cubic polynomial in Eq.~\eqref{eq:cubic} and the ellipse
\begin{equation}
 Q_D(\lambda,\beta)=100\,Q(\lambda,\beta)<1 ,
 \label{eq:supp-config-D}
\end{equation}
with $Q$ from Eq.~\eqref{eq:main-q-tor}.  This configuration could be used
to benchmark fixed-boundary VMEC calculations.

\end{document}